\documentclass[12pt]{article}
\usepackage{graphics}
\newcommand{\pubnum}{473}
\newcommand{\data}{September 1999}

\newcommand{\sq}{\ensuremath{\tilde q}} 
\newcommand{\su}{\ensuremath{\tilde u}} 
\newcommand{\sd}{\ensuremath{\tilde d}} 
\newcommand{\stopp}{\ensuremath{\tilde t}} 
\newcommand{\cplus}{\ensuremath{\chi^+}} 
\newcommand{\gsim}{\mbox{ \raisebox{-4pt}{${\stackrel{\textstyle >}{\sim}}$} }}\newcommand{\lsim}{\mbox{ \raisebox{-4pt}{${\stackrel{\textstyle <}{\sim}}$} }}\newcommand{\mA}{\ensuremath{M_{A^0}}} 
\newcommand{\tch}{{t\rightarrow c\,h}} 
\newcommand{\pl}{P_L} 
\newcommand{\pr}{P_R} 
\newcommand{\Amaidu}{\ensuremath{A_{-a i}^{(d,u)}}}
\newcommand{\Apaidu}{\ensuremath{A_{+a i}^{(d,u)}}}

\begin{document}
\thispagestyle{empty} \hbox{
\vrule height0pt width5in
\vbox{\hbox{\rm
UAB-FT-\pubnum
}\break
\hbox{KA-TP-15-1999\hfill}
\break\hbox{hep-ph/9909503\hfill}
\break\hbox{\data\hfill}
\hrule height.1cm width0pt}
} \vspace{3mm}

\begin{center}
{\large \textbf{Top quark decays into neutral Higgs bosons and gluon in the
MSSM}}\footnote{Updated talk presented in: \emph{International Workshop on
Linear Colliders}, Sitges, April 28-May 5, 1999; to appear in the proceedings.}

\bigskip\vskip8mm

{\large Jaume Guasch}$^{a}${\large , Joan Sol\`{a}}$^{b}$ \vskip3mm

\medskip

$^{a}$\textsl{Institut f\"{u}r Theorestische Physik, Universit\"{a}t Karlsruhe,}

\textsl{\ D-76128 Karlsruhe, Germany}

$^{b}$\textsl{Grup de Fisica Te\`{o}rica and Institut de Fisica d'Altes
Energies, }

\textsl{Universitat Aut\`{o}noma de Barcelona, E-08193, Bellaterra, Barcelona,
Catalonia, Spain}

\medskip
\end{center}

\vspace{0.3cm}

\begin{center}
\textbf{ABSTRACT}
\end{center}

\begin{quotation}
Flavour changing neutral decays of the top quark are know to be extremely
suppressed in the SM. This is especially so for the top quark decay into the
SM Higgs boson, whose rate is less than $10^{-13}$. However, it turns out that
the decay modes $t\rightarrow c\,h$, with $h\equiv h^{0},H^{0},A^{0}$ any of
the MSSM neutral Higgs bosons, can be much more gifted. In particular, the
rate into the lightest CP-even Higgs boson $h^{0}$ -- which is an accessible
decay mode across the whole MSSM parameter space -- is generally higher than
the decay into glue, $t\rightarrow c\,g$, even though the latter can also be
considerably augmented in the MSSM. Our general conclusion is that the Higgs
channel $t\rightarrow c\,h^{0}$ can be the ``gold-plated'' top quark FCNC
decay mode in the MSSM with rates that can reach $10^{-4}$, i.e. already at
the visible level for both LHC and LC.
\end{quotation}

\newpage

\section{Introduction}

\label{sec:intro}

At the tree-level there are no FCNC processes in the SM, and at one-loop they
are induced by charged-current interactions, which are GIM-suppressed. In
particular, FCNC decays of the top quark into gauge bosons ($t\rightarrow
c\,V$;$\;V\equiv\gamma,Z,g$) are very unlikely, with maximum rates of order
$10^{-12}$ for the photon, slightly above $10^{-13}$ for the $Z$-boson, and at
most $10^{-10}$ for the gluon channel\,\cite{GEilam,FCNCSM}. Even more
dramatic is the situation with the top quark decay into the SM Higgs boson,
$t\rightarrow c\,H_{SM}$, which has recently been recognized to be much more
disfavored than originally thought\thinspace\cite{GEilam}: $BR(t\rightarrow
c\,H_{SM})=1\cdot10^{-13}-4\cdot10^{-15}$ $(M_{Z}\leq M_{H}\leq2\;M_{W}%
)$\thinspace\cite{Mele}. This rate is far out of the range to be covered by
any presently conceivable high luminosity machine. On the other hand, the
highest SM rate, namely that of $t\rightarrow c\,g$, is still $5$ orders of
magnitude below the feasible experimental possibilities at the LHC. All in all
detection of FCNC decays of the top quark at visible levels (viz.
$\,BR(t\rightarrow c\,X)\gsim10^{-5}-10^{-4}$) by any of the future high
luminosity colliders round the corner (especially LHC and LC) seems doomed to
failure in the absence of new physics.

Thus the possibility of large enhancements of some FCNC rates up to visible
levels, particularly within the context of general two-Higgs-doublet models
($2${HDM) and especially in the Minimal Supersymmetric Standard Model
(}MSSM)\,\cite{Hunter}, should be very welcome. In the LHC, for example, the
production of top quark pairs will be very high: $\sigma(t\overline
{t})=800\;pb$ -- {roughly two orders of magnitude larger than that of the
Tevatron II at }$\sqrt{s}=2\,TeV$. In the so-called low-luminosity phase
($10^{33}\,cm^{-2}s^{-1}$) of the LHC one expects ten million $t\,\bar{t}%
$-pairs per year\,\cite{Gianotti}. And this number will be augmented by one
order of magnitude in the high-luminosity phase ($10^{34}\,cm^{-2}s^{-1}$). As
for a future $e^{+}e^{-}$ linear collider running at e.g. $\sqrt{s}=500\;GeV$,
one has a smaller cross-section $\sigma(t\bar{t})=650\;fb$ but a higher
luminosity factor ranging from $5\times10^{33}\,cm^{-2}s^{-1}$ to
$5\times10^{34}\,cm^{-2}s^{-1}$ and of course a much cleaner
environment\,\cite{LCSitges}. Thus, {with datasets from LHC and LC increasing
to several }$100\,fb^{-1}/${year in the high-luminosity phase, one should be
able to collect enough statistics (perhaps some few hundred to few thousand
events) from the combined output of these machines enabling us to perform an
efficient study of rare top quark decays beyond the SM. A detailed
presentation of this work is given in Ref.\cite{NP}.}

\section{Relevant decays and Lagrangians in the MSSM}

The Higgs channels in the MSSM are especially promising. They comprise the top
quark decays {into the two CP-even (``scalar'') states and the CP-odd
(``pseudoscalar'') state of the Higgs sector of the MSSM}\,\cite{Hunter},
\begin{equation}
t\rightarrow c\,h\ \ \ \ (h=h^{0},H^{0},A^{0}). \label{tch}%
\end{equation}
{Worth emphasizing} is the fact that in the MSSM (in contrast to the SM or the
unconstrained $2${HDM) at least one of these decays (viz. }$t\rightarrow
c\,h^{0}${) is \textsl{always} possible, for in the MSSM there is an upper
bound on the mass of the lightest CP-even Higgs boson, }$M_{h^{0}}%
\lsim135\,GeV${\,\cite{HollWeig}, which is below the top quark mass. Moreover,
for a sufficiently light pseudoscalar mass, }$M_{A^{0}}<m_{t},$ {all three
decays (\ref{tch}) are in principle possible, if the SUSY masses are not that
high so as to induce too large positive corrections to }$M_{H^{0}}$.
Preliminary studies of the decays (\ref{tch}) were presented in\,\cite{YangLi}%
, but they did not include the one-loop Higgs mass relations and moreover they
missed some of the most relevant effects.

We report here also on the MSSM rate of the most favored FCNC top quark decay
in the SM, namely the top decay into glue
\begin{equation}
t\rightarrow c\;g\,. \label{tcglue}%
\end{equation}

This channel has been extensively studied in the literature, both in the
SM\,\cite{GEilam,FCNCSM} and in the MSSM\,\cite{Divitiis,Koenig}. However, we
include a simultaneous numerical analysis of it in order to use it as a
fiducial mode with which to better assess the size of our results on the Higgs
channels (\ref{tch}). As for the remaining FCNC decay modes, namely
$t\rightarrow c\,(\gamma,Z),$ they have also been evaluated in the MSSM, but
in spite of some enhancements the general conclusion is that they are
hopeless\,\cite{Divitiis,tcV}.

%
%
\begin{figure}[ptb]
\begin{center}
\resizebox{9.5cm}{!}{\includegraphics{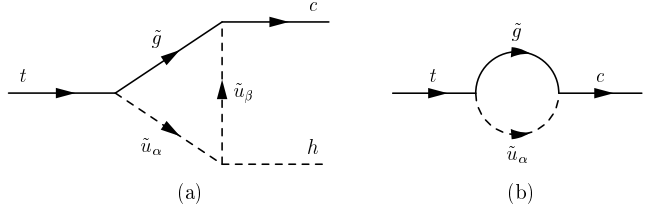}}
\end{center}
\caption{One-loop SUSY-QCD diagrams for the decay $\tch$: \textbf{(a)} vertex
diagram, \textbf{(b)} mixed $t-c$ self-energy. $\su_{\{\alpha,\beta\}}$ stand
for mass-eigenstate up-type squarks of any generation.}%
\label{diag:fcncQCDvertex}%
\end{figure}

The basic interactions responsible for the enhancements lie both in the QCD
sector and in the electroweak (EW) sector of the MSSM. The SUSY-QCD
interactions are described by the following Lagrangian, in the mass-eigenstate
basis:
\begin{eqnarray}
\mathcal{L}_{\mathrm{SUSY-QCD}}  &  =&-\frac{g_{s}}{\sqrt{2}}\,\bar{\psi}%
_{c}^{{\tilde{g}}}\left[  R_{5\alpha}^{\ast}\,P_{L}-R_{6\alpha}^{\ast}%
\,P_{R}\right]  {\tilde{q}}_{\alpha,i}^{\ast}\,\lambda_{ij}^{c}\,t_{j}%
\nonumber\\
&  -&\frac{g_{s}}{\sqrt{2}}\,\bar{\psi}_{c}^{{\tilde{g}}}\left[  R_{3\alpha
}^{\ast}\,P_{L}-R_{4\alpha}^{\ast}\,P_{R}\right]  {\tilde{q}}_{\alpha,i}%
^{\ast}\,\lambda_{ij}^{c}\,c_{j}\nonumber\\
&  -&\frac{g_{s}}{\sqrt{2}}\,\bar{\psi}_{c}^{{\tilde{g}}}\left[  R_{1\alpha
}^{\ast}\,P_{L}-R_{2\alpha}^{\ast}\,P_{R}\right]  {\tilde{q}}_{\alpha,i}%
^{\ast}\,\lambda_{ij}^{c}\,u_{j}+\mbox{ h.c.}\,\,. \label{SUSYQCD}%
\end{eqnarray}
$\psi_{c}^{{\tilde{g}}}$ stands for the gluino spinor, $\lambda^{c}$ are the
$SU(3)_{c}$ Gell-Mann matrices and $P_{L,R}$ the chiral projector operators.
The $6\times6$ rotation matrices $R^{(q)}$ are needed to diagonalize the
squark mass matrices in (flavor)$\times$(chiral) space as follows (notation as
in\,\cite{NP}):
\begin{eqnarray}
\sq_{\alpha}^{\prime}  &  =&\sum_{\beta}R_{\alpha\beta}^{(q)}\sq_{\beta}^{{}%
}\,\,,\nonumber\label{eq:definicioR6gen}\\
R^{(q)\dagger}\mathcal{M}_{\sq}^{2}R  &  =&\mathcal{M}_{\sq D}^{2}%
=\mathrm{diag}\{m_{\sq_{1}}^{2},\ldots,m_{\sq_{6}}^{2}\}\,\,\,\,(q\equiv
u,\,d)\,\,,
\end{eqnarray}
where $\mathcal{M}_{(\su,\sd)}^{2}$ is the square mass matrix for squarks in
the EW basis ($\sq_{\alpha}^{\prime}$), with indices $\alpha=1,2,3,\ldots
,6\equiv\su_{L},\su_{R},\tilde{c}_{L},\ldots,\stopp_{R}$ for up-type squarks,
and a similar assignment for down-type squarks. The intergenerational mixing
terms leading to gluino-mediated FCNC couplings lie in the off-diagonal
entries of the mass matrices. However, in order to prevent the number of
parameters from being too large, we have allowed (symmetric) mixing mass terms
only for the left-handed (LH) squarks. This simplification is often used in
the MSSM and it is justified by Renormalization Group (RG) analysis\thinspace
\cite{duncan}. Following this practice, we introduce flavor-mixing
coefficients $\delta_{ij}$ in the LL block of the $6\times6$ squark mass
matrix (namely the one involving only LH fields of any flavor) as follows:
\begin{equation}
(M_{LL}^{2})_{ij}=m_{ij}^{2}\equiv\delta_{ij}\,m_{i}\,m_{j}\ \ (i\neq j)\,\,,
\label{eq:defdelta}%
\end{equation}
%
%
\begin{figure}[ptb]
\begin{center}
\resizebox{9.5cm}{!}{\includegraphics{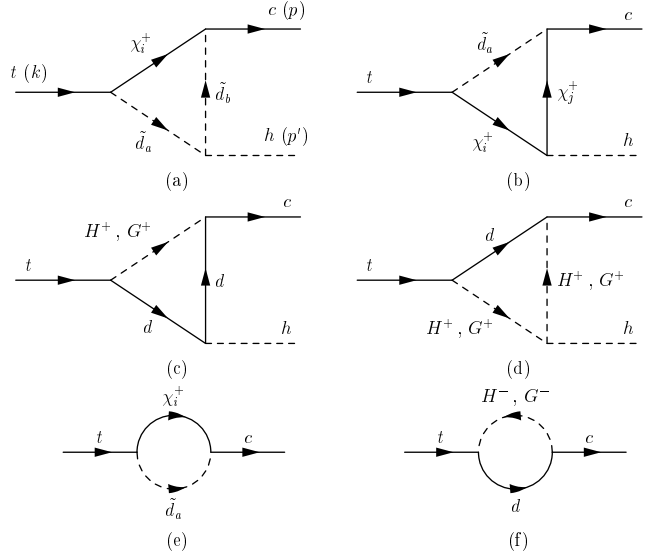}}
\end{center}
\caption{One-loop SUSY-EW diagrams for the decay $\tch$ $(h=h^{0},H^{0}%
,A^{0})$. Here $d$ ($\sd_{\{a,b\}}$) represent mass-eigenstate down type
quarks (squarks) of any generation.}%
\label{diag:fcncEWvertex}%
\end{figure}where $m_{i}$ is the mass of the left-handed $i$th squark, and
$m_{ij}^{2}$ is the mixing mass matrix element between generations $i$ and
$j$. Therefore, if the coefficients $\delta_{ij}$ are non-vanishing, for some
$i\neq j$, then the structure of the diagonalizing matrices $R^{(q)}$ defined
above must necessarily lead to gluino-mediated tree-level FCNC between quarks
and squarks in the SUSY-QCD Lagrangian (\ref{SUSYQCD}). This scenario can be
generalized by further introducing FCNC interactions on the right-handed (RH)
block of the mass matrix, but this hypothesis entails an unnatural departure
from the RG expectations and moreover it is not necessary to achieve the
visible rates\,\cite{NP}.

On the other hand, the leading EW interactions appear through the possibility
of large Yukawa couplings (normalized with respect to the $SU(2)_{L}$ gauge
coupling):%
\begin{equation}
\lambda_{u}\equiv{\frac{h_{u}}{g}=\frac{m_{u}}{\sqrt{2}\,M_{W}\,\sin{\beta}}%
}\;\;\;\;\;,\;\;\;\;\;\lambda_{d}\equiv{\frac{h_{d}}{g}}={\frac{m_{d}}%
{\sqrt{2}\,M_{W}\,\cos{\beta}}}\,\,. \label{eq:Yukawas}%
\end{equation}
For the computation of the EW effects we work in the approximation that the
squark mass matrices diagonalize whith the same matrix elements than the
quarks, the standard CKM-matrix. Although we will not spell out here the
structure of the various relevant interactions in the EW sector, let us for
the sake of illustration to quote the interaction Lagrangian involving
charginos, quarks and squarks in the mass-eigenstate basis (notation as
in\,\cite{GSZP}):
\begin{equation}
{\cal L}_{u\,\sd\,\cplus}=-g\,\sum_{\stackrel{\scriptstyle d=d,s,b}%
{u=u,c,t}} V_{ud}\,\sd_{a}^{\ast}\,\bar{\psi}_{i}^{+}\left(  \Apaidu
\,\pl+\Amaidu\,\pr\right)  \,u+\mbox{ h.c.}\,\,, \label{eq:Lsqsqc}%
\end{equation}
$V_{ud}$ is the standard CKM-matrix element, and the coupling matrices are
defined as
\begin{equation}
\Apaidu=R_{1a}^{(d)\ast}V_{i1}^{\ast}-\lambda_{d}\,R_{2a}^{(d)\ast}%
V_{i2}^{\ast}\,\,,\,\,\Amaidu=-R_{1a}^{(d)\ast}\lambda_{u}U_{i2}\,\,,
\end{equation}
with $V_{ij}$ and $U_{ij}$ the rotation matrices of the chargino
sector\,\cite{GSZP}. Here $\lambda_{u}$, $\lambda_{d}$ are the up-like and
down-like Yukawa couplings (\ref{eq:Yukawas}). Its significance is determined
by the value of the parameter $\tan\beta=v_{2}/v_{1}$\,\cite{Hunter}. At large
$\tan\beta$ one expects that the EW interactions may play here a significant
role as in the case of top quark decays into charged Higgs\,\cite{CGGJS}. From
these interaction Lagrangians, the Feynman diagrams responsible for the FCNC
decays (\ref{tch}) can be depicted in Fig.1 (SUSY-QCD effects) and Fig. 2
(SUSY-EW). One may draw similar diagrams for the decay (\ref{tcglue}).

\section{Numerical Analysis and Discussion}

We use the fiducial ratios
\begin{equation}
B({t\rightarrow c\,h})\equiv\frac{\Gamma(t\rightarrow c\,h)}{\Gamma
(t\rightarrow b\,W^{+})}\,,\;B({t\rightarrow c\,g})\equiv\frac{\Gamma
(t\rightarrow c\,g)}{\Gamma(t\rightarrow b\,W^{+})} \label{eq:defbr}%
\end{equation}
to carry out our numerical analysis.

These ratios are not the total branching fractions $BR({t\rightarrow c\,X})$
for the FCNC decay modes under study $(X=h,g)$, as there are many other
channels that should be added up to the denominator of (\ref{eq:defbr}) in the
MSSM, if kinematically allowed. However, for better comparison with previous
analyses of FCNC top quark decays, the ratios (\ref{eq:defbr}) should suffice
to assess the experimental viability of the FCNC decays under consideration.

After an exhaustive scanning of the MSSM parameter space we find the following
results\,\cite{NP}. The most important source of FCNC enhancement comes from
the SUSY-QCD sector, whose numerical effect hinges on the values of the
flavour-mixing coefficients $\delta_{ij}$ in eq.(\ref{eq:defdelta}). The
present experimental bounds on these quantities depend on the value of the
squark masses and read as follows\,\cite{gabbiani}:
\begin{eqnarray}
|\delta_{12}|  &  <&.1\,\sqrt{m_{\tilde{u}}\,m_{\tilde{c}}}/{500{\,GeV}%
}\,\,,\nonumber\\
|\delta_{13}|  &  <&.098\,\sqrt{m_{\tilde{u}}\,m_{\tilde{t}}}/{500{\,GeV}%
}\,\,,\nonumber\\
|\delta_{23}|  &  <&8.2\,m_{\tilde{c}}\,m_{\tilde{t}}/{(500{\,GeV})^{2}}\,\,\,.
\label{eq:limdelta}%
\end{eqnarray}
In using these bounds we make use of $SU(2)$ gauge invariance to transfer the
experimental information known from the down-quark sector (for example from
{$BR(b\rightarrow s\;\gamma)$}, where the bound on $\delta_{23}$ is obtained)
to the up-quark sector.

With a SUSY mass spectrum of a few hundred ${\,GeV}$, we find that the
different contributions to the Higgs channels are typically of the order
\begin{eqnarray}
B^{\mathrm{SUSY-EW}}({t\rightarrow c\,h})  &  \simeq&10^{-8}%
\,\,,\nonumber\label{eq:confinal}\\
B^{\mathrm{SUSY-QCD}}({t\rightarrow c\,h})  &  \simeq&10^{-5}\,\,.
\end{eqnarray}
However, by stretching out a bit more the range of parameters one can reach
({for some of the decays})
\begin{eqnarray}
B^{\mathrm{SUSY-EW}}({t\rightarrow c\,h})  &  \simeq&1\times10^{-6}%
\,\,,\nonumber\\
B^{\mathrm{SUSY-QCD}}({t\rightarrow c\,h})  &  \simeq&5\times10^{-4}.
\label{corner}%
\end{eqnarray}

The difference of at least two orders of magnitude between the SUSY-EW and
SUSY-QCD contributions makes unnecessary to compute the interference terms
between the two sets of amplitudes.

{We have obtained the maximum rates (\ref{corner}) from a general search in
the MSSM parameter space within the }$1\,TeV$ {mass region and respecting the
experimental constraints, in particular the relations (\ref{eq:limdelta}). }In
Fig.\thinspace3 we present the maximum values that can be reached by
$B(t\rightarrow c\,X)\,$ for each of the processes presented. In
Figs.\thinspace3a and~3b we show the maximized $B(t\rightarrow c\,h)$ as a
function of the pseudoscalar Higgs boson mass by taking into account only the
SUSY-EW contributions and the SUSY-QCD contributions respectively. Perhaps the
most noticeable result is that the decay into the lightest MSSM Higgs boson
($t\rightarrow c\,h^{0}$) is the one that can be maximally enhanced and
reaching values of order $B(t\rightarrow c\,h^{0})\sim10^{-4}$ that stay
fairly stable all over the parameter space. The reason for this dominance is
that the decay $t\rightarrow c\,h^{0}$ is the one which is more sensitive to
the trilinear coupling $A_{t}$, a parameter whose natural range reaches up to
about $1\,TeV$.
%
%
\begin{figure}[ptb]
\begin{center}%
\begin{tabular}
[c]{cc}%
\resizebox{5cm}{!}{\includegraphics{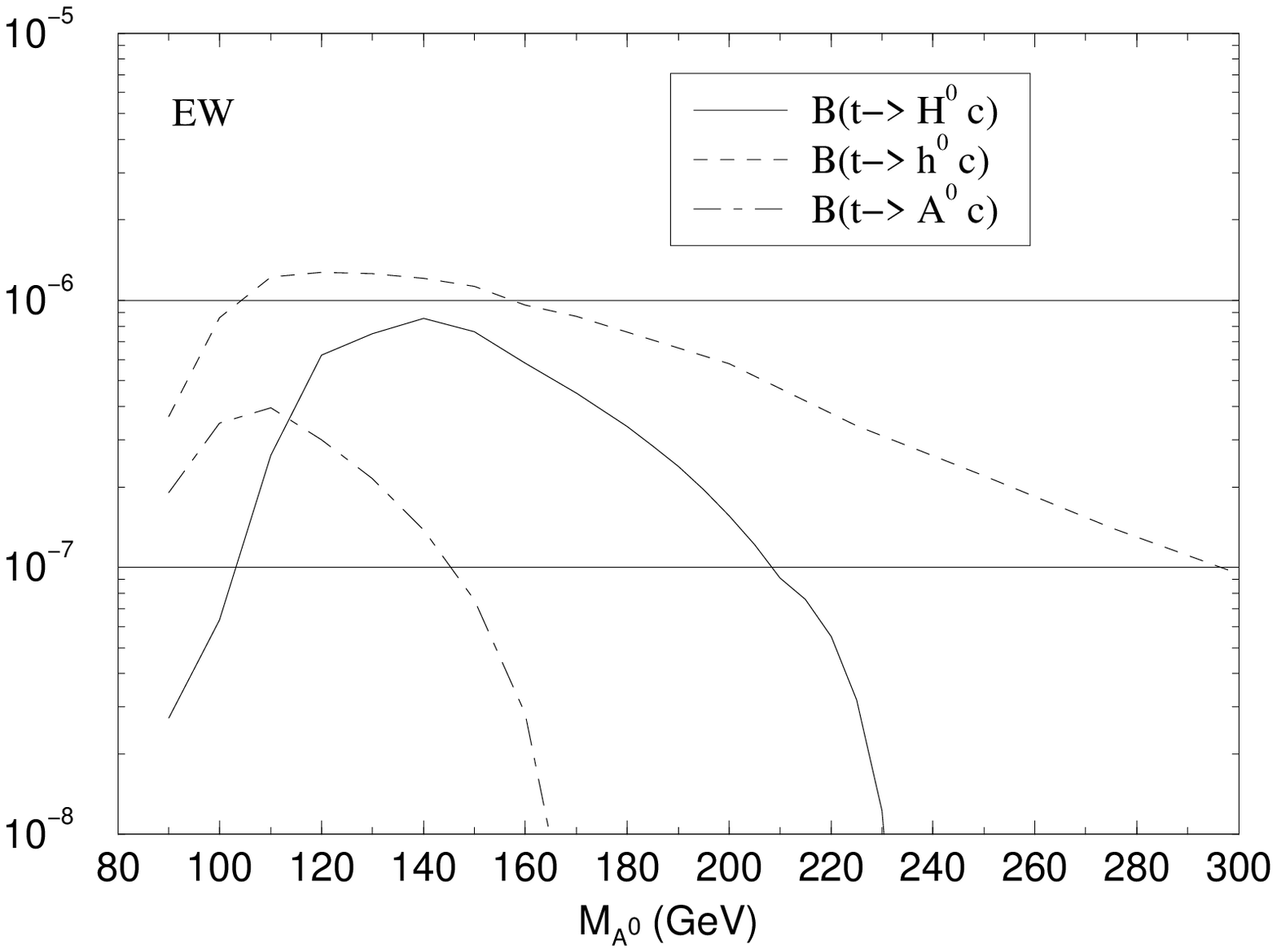}} & \resizebox{5cm}%
{!}{\includegraphics{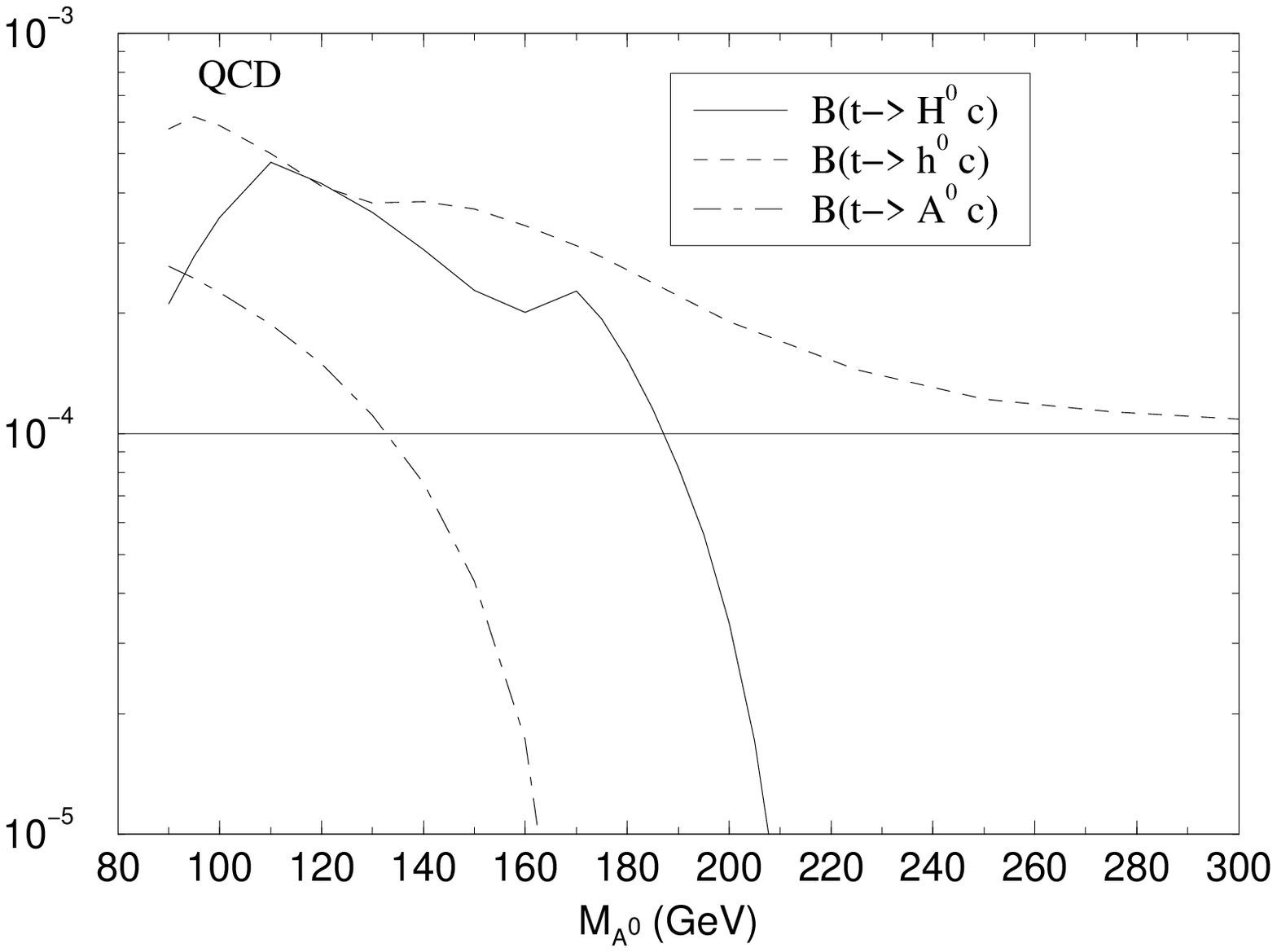}}\\
(a) & (b)\\
\multicolumn{2}{c}{\resizebox{5cm}{!}{\includegraphics{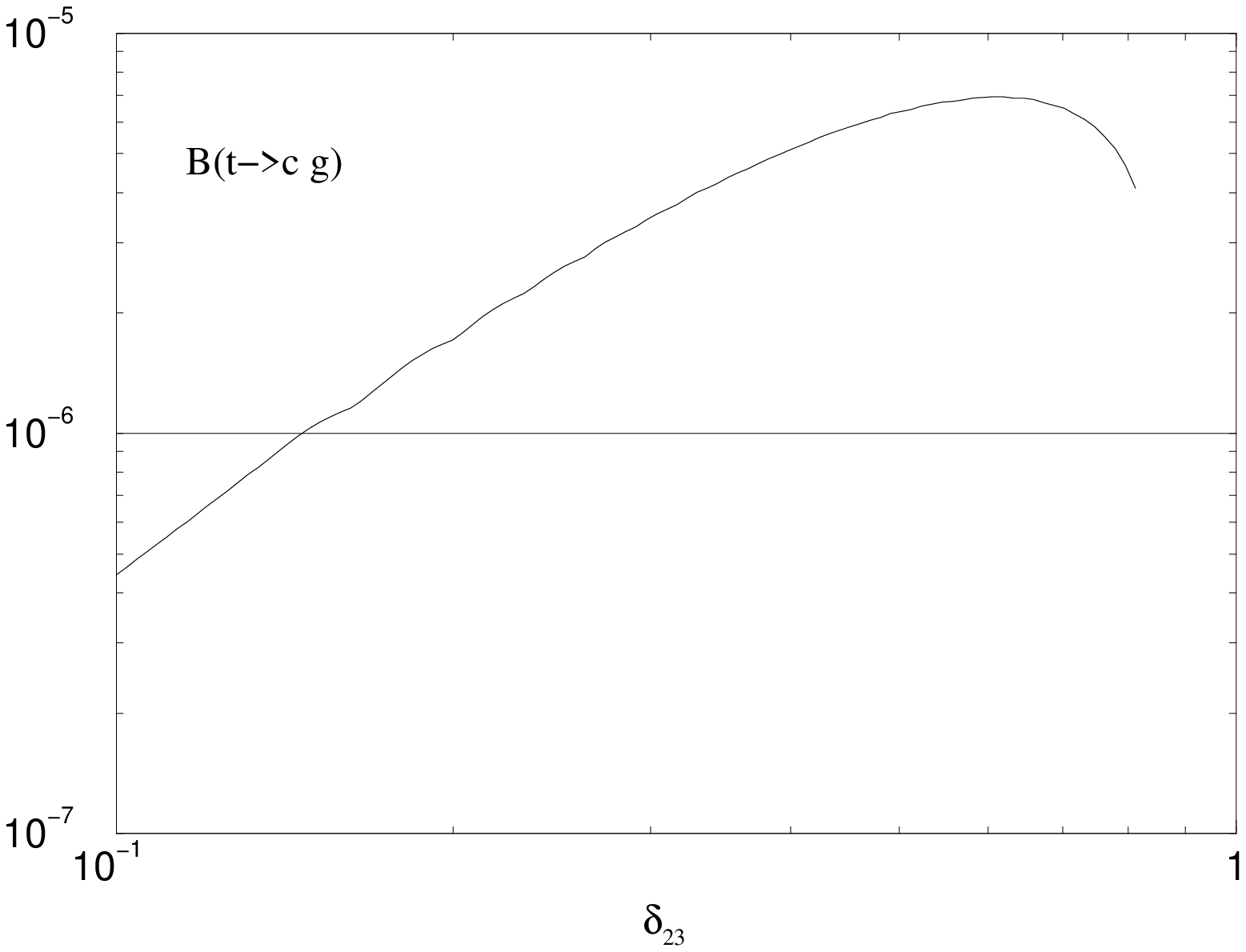}%
}}\\
\multicolumn{2}{c}{(c)}%
\end{tabular}
\end{center}
\caption{\textbf{(a) }Maximum value of $B(t\rightarrow c\,h)$, obtained by
taking into account only the SUSY-EW contributions, as a function of $\mA$ ;
\textbf{(b)} as in (a) but taking into account only the SUSY-QCD
contributions; and \textbf{(c)} maximum value of $B(t\rightarrow c\,g)$ as a
function of the intergenerational mixing parameter $\delta_{23}$ in the LH
sector. In all cases the scanning for the rest of parameters of the MSSM has
been performed within the phenomenologically allowed region.}%
\label{fig:maxims}%
\end{figure}

{We have also studied the optimal FCNC rates of the gluon channel in the
MSSM.} In Fig.\thinspace3c we have plotted the maximum value of
$B(t\rightarrow c\,g)$ as a function of $\delta_{23}$ after scanning for the
rest of the MSSM mass parameters within the $1\,TeV$ range. {{Under the
RG-based assumption of only mixing in the LH sector}, one has }%
$B^{\mathrm{SUSY-QCD}}({t\rightarrow c\,g})\lsim\,10^{-5}$. {Our results on
this channel are compatible with the recent analysis of Ref.\,\cite{Divitiis}
in contrast to that of Ref.\thinspace\cite{Koenig}}.

{To assess the discovery reach of the FCNC top quark decays in the next
generation of accelerators} we take as a guide the estimations that have been
made for gauge boson final estates\,\cite{limits}. Using the information
mentioned in Sect.\thinspace\ref{sec:intro} and assuming that all the FCNC
decays $t\rightarrow c\,X$ ($X=V,h$) can be treated similarly, we roughly
estimate the following sensitivities for $100\;fb^{-1}$ of integrated
luminosity:
\begin{eqnarray}
\mathrm{\mathbf{LHC:}}B(t\rightarrow c\,X)  &  \gsim&5\times10^{-5}\nonumber\\
\mathrm{\mathbf{LC:}}B(t\rightarrow c\,X)  &  \gsim&5\times10^{-4}\nonumber\\
\mathrm{\mathbf{TEV33:}}B(t\rightarrow c\,X)  &  \gsim&5\times10^{-3}\,\,\,.
\label{sensitiv}%
\end{eqnarray}
Therefore, LHC seems to be the most suitable collider where to test this kind
of phenomena. The LC is limited by statistics (due to much smaller top quark
cross-section) but {in compensation} every collected event is clear-cut. So
this machine could eventually be of much help, especially if we take into
account that it could deliver $500\;fb^{-1}$ {per year}\thinspace
\cite{LCSitges}.{ The situation with the Tevatron, however, is much more
gloomy, since the required FCNC rates of} $\sim10^{-3}$ {cannot be attained
unless we artificially search in some remote (unnatural) corner of the
parameter space.}

To conclude, the FCNC decays of the top quark are such rare events in the SM
(especially the top quark decay into the Higgs boson) that their observation
at detectable levels should be interpreted as an extremely robust indication
of new physics. The effective tagging of FCNC top quark decays in the major
accelerators round the corner could well be a first step into discovering
SUSY. From our analysis we find the following MSSM maximum rates for the most
favorable modes:
\begin{equation}
5\times10^{-6}\lsim B(t\rightarrow c\,g)_{\max}<B(t\rightarrow c\,h^{0}%
)_{\max}\lsim5\times10^{-4}.\label{maxs}%
\end{equation}
{In both decays the dominant effects come from SUSY-QCD. However, it should
not be undervalued the fact that the maximum electroweak rates for
}$t\rightarrow c\,h$ {can reach the }$10^{-6}$ {level, i.e. on the verge of
being detectable. }

\section*{Acknowledgments}

This work has been partially financed by CICYT under project No. AEN95-0882
and by the Deutsche Forschungsgemeinschaft.

\end{document}